\documentclass[pra,superscriptaddress,showpacs,preprint]{revtex4}
\usepackage[latin1]{inputenc}

\newcommand{\EVRY}{D\'epartement de Physique et Mod\'elisation,
Universit\'e d'Evry Val d'Essonne\\
Boulevard F. Mitterrand, 91025 Evry, France}
\newcommand{\LKB}{Laboratoire Kastler Brossel, Universit\'e Pierre et Marie Curie\\
T12, Case 74, 4 place Jussieu, 75252 Paris, France}

\begin{document}
\title{Vibrational spectroscopy of H$_2^+$: \\
precise evaluation of the Zeeman effect }

\author{Jean-Philippe Karr}
\email{karr@spectro.jussieu.fr}
\affiliation{\LKB}
\affiliation{\EVRY}

\author{Vladimir I.~Korobov}
\affiliation{Joint Institute for Nuclear Research, 141980, Dubna, Russia}

\author{Laurent Hilico}
\affiliation{\LKB}
\affiliation{\EVRY}

\date{\today}
\begin{abstract}
We present an accurate computation of the $g$-factors of the hyperfine states of the hydrogen molecular ion H$_2^+$. The results are in good agreement with previous experiments, and can be tested further by rf spectroscopy. Their implication for high-precision two-photon vibrational spectroscopy of H$_2^+$ is also discussed. It is found that the most intense hyperfine components of two-photon lines benefit from a very small Zeeman splitting.
\end{abstract}
\maketitle
\section{Introduction}
In~\cite{paper1}, we have studied the spectrum of two-photon ro-vibrational transitions in the hydrogen molecular ion H$_2^+$. The precise measurement (using a Doppler-free excitation geometry) of the frequency of such transitions in a rf trap constitutes a promising new method for determination of the electron-to-proton mass ratio $m_e/m_p$~\cite{karr2007,koelemeij2007}. Our estimate of transition rates with present-day experimental parameters has shown the feasibility of such an experiment. In order to assess the metrological merits of the method, it is essential to evaluate the order of magnitude of systematic effects; if one of them turns out to be a limiting factor, it is desirable to calculate it precisely in order to subtract its effects from the experimental data.

Among the systematic effects, the ac Stark shifts due to the optical fields, blackbody radiation and rf trap potential are expected to be small, due to the small dynamic polarizability of the H$_2^+$ ion~\cite{hilico2001}. The second-order Doppler effect results in a shift and broadening of the line by about 10 kHz in a typical Paul trap, but can be reduced by sympathetic ion cooling~\cite{roth2005}. The Zeeman shift is one of the most important effects remaining to be investigated, especially if circular polarization is used, resulting in the selection rule $\Delta M_J = \pm$2 for two-photon transitions. In the low magnetic field regime, the hyperfine structure has to be taken into account. The $g$-factors of the hyperfine levels have been calculated long ago~\cite{mizushima1960} but with an accuracy limited by an imperfect knowledge of the hyperfine structure. A measurement of $g$-factors ratios has been used by Richardson et al. to extract improved values of the hyperfine Hamiltonian coefficients~\cite{richardson1968}. We have recently determined these coefficients {\em ab initio} with an improved relative accuracy of $\mathcal{O}(\alpha^2)$, corresponding to the limit of the Breit-Pauli Hamiltonian~\cite{korobov2006}. The main aim of the present work is to calculate the $g$-factors with much better accuracy than obtained so far. This high accuracy is, in fact, not needed for optical spectroscopy experiments, where the magnetic field is usually not controlled very precisely, so that the uncertainty on the Zeeman shift will be limited by the uncertainty on the magnetic field and its eventual fluctuations. However, we think it is worthwhile to present these precise values, because they can be tested in rf spectroscopy experiments of the same type as described in~\cite{richardson1968}, which would provide a good test of hyperfine structure calculations.

\section{Zeeman Hamiltonian}
Neglecting relativistic and radiative corrections, the linear part of the Hamiltonian describing interaction of a H$_2^+$ ion with a magnetic field is given by
\begin{equation}
H_Z = g_e \mu_B \; \mathbf{S}_e \cdot \mathbf{B} - g_p \mu_p \; \mathbf{I} \cdot \mathbf{B} + \mu_B \; \mathbf{L}_e \cdot \mathbf{B} - \mu_p \; \left( \mathbf{L}_1 \!+\! \mathbf{L}_2 \right) \cdot \mathbf{B}, \label{hamiltonian}
\end{equation}
where $g_e$ and $g_p$ are respectively the electron and proton $g$-factors, $\mu_B = e/2 m_e$ is the Bohr magneton, and $\mathbf{L}_e, \mathbf{L}_1, \mathbf{L}_2$ are the orbital momenta of the electron and both protons in the center-of-mass frame. $\mathbf{S}_e$ and $\mathbf{I}$ are the electron and nuclear spins (see Ref.~\cite{paper1}). The magnetic field is assumed to be oriented along $Oz$. In the ground electronic state 1s$\sigma_g$, the main contribution to the $g$-factor comes from the first term, i.e. the electron magnetic moment, while the other terms are about a factor of 1000 smaller. For the third term this can be understood by noting that the rotation velocities of the electron and protons are of the same order; hence the terms in $\mathbf{L}_e$ and $\mathbf{L}_1 \!+\! \mathbf{L}_2$ are of the same order.

In low magnetic fields, the hyperfine structure has to be taken into account, and the Zeeman Hamiltonian can be written using the Landé factor (or $g$-factor) of the hyperfine level under consideration:
\begin{equation}
\tilde{H}_Z = g_J \mu_B\; \mathbf{J} \cdot \mathbf{B}
\end{equation}
where $\mathbf{J}$ is the total angular momentum; note that the scalar $g_J$ becomes a tensor if relativistic and radiative corrections are taken into account, in contradistinction with the atomic case, due to the lack of central symmetry of the potential~\cite{hegstrom1979}.

Let us briefly recall the structure of hyperfine levels (for more details, see~\cite{paper1}). For even values of $L$, the total nuclear spin $I$ is zero the total angular momentum is $\mathbf{J} = \mathbf{L} + \mathbf{S}_e$, and each ro-vibrational level $(v,L)$ is split into two hyperfine levels $(v,L,J)$ with $J\!=\!L\!\pm\!1/2$. For odd values of $L$, $I$ is equal to one and the coupling scheme is as follows: the total spin $\mathbf{F} = \mathbf{S}_e + \mathbf{I}$ ($F\!=\!1/2$ or $3/2$) which is not an exact quantum number, and the total angular momentum $\mathbf{J} = \mathbf{L} + \mathbf{F}$. $J$ can take the values $L \pm 3/2$, $L \pm 1/2$. If $J = L \pm 3/2$, the hyperfine state is a pure $F\!=\!3/2$ state $\left|v,L,S_e\!=\!1/2,I\!=\!1,F\!=\!3/2,J\!=\!L\!\pm\!3/2,M_J\right\rangle$. If $J = L \pm 1/2$, the hyperfine state is a linear combination of $F\!=\!1/2$ and $F\!=\!3/2$ states:
\begin{equation}
\textstyle
\bigl|\,v,L,S_e,I,\tilde{F}, J\!=\!L\!\pm\!1/2,M_J \bigr\rangle \>\equiv\>
C^{\pm}_1 \,\bigl|\,v,L, 1/2 ,1, 1/2 ,L\!\pm\!1/2,M_J \bigr\rangle
\>+\>
C^{\pm}_3 \,\bigl|\,v,L,1/2,1,3/2,L\!\pm\!1/2,M_J \bigr\rangle.
\label{eq-coefCi1}
\end{equation}
In the following, we evaluate the contribution of each term in equation~(\ref{hamiltonian}) to the $g$-factor of the pure states $\left| v,L,S_e,I,F,J,M_J \right\rangle$, noted as $g_J(v,L,F,J)$ (if $L$ is even, these notations are replaced by $\left| v,L,S_e,J \right\rangle$ and $g_J(v,L,J)$, respectively). The effect of state mixing will be addressed in Sec.~\ref{section-mixing}. The $g$-factor is divided into three contributions:
\begin{equation}
g_J(v,L,F,J) = g_1 (L,F,J) + g_2 (L,F,J) + g_3 (v,L,F,J)
\end{equation}
where $g_1$ is the contribution from the first term in equation~(\ref{hamiltonian}), $g_2$ is the contribution from the second term, and $g_3$ the contribution from the last two terms. Note that the first two quantities do not depend on the vibrational quantum number $v$, because the Hamiltonian only involves spin operators. In contradistinction, the last term contains orbital momentum operators acting on the orbital wave function, which introduces a slight dependence on $v$.

The standard angular algebra procedures used below can be found in many textbooks, e.g.~\cite{messiah}.

\section{Contribution of the electron spin}
Here, we evaluate the contribution to the $g$-factor coming from the first term in equation~(\ref{hamiltonian}). The Zeeman shift of a given hyperfine level $\left| v,L,S_e,I,F,J,M_J \right\rangle$ due to this term is given by
\begin{equation}
\Delta E_1 = g_e \mu_B B \left\langle v,L,S_e,I,F,J,M_J | S_e^z | v,L,S_e,I,F,J,M_J \right\rangle.
\end{equation}
From now on the dependence in $v$ will not be noted, since $S_e^z$ does not act on the orbital wave function. Application of the Wigner-Eckart theorem to the vector operator $\mathbf{S}_e$ yields
\begin{equation}
\begin{array}{@{}rcl}
\left\langle L,S_e,I,F,J,M_J | S_e^z | L,S_e,I,F,J,M_J \right\rangle
&=&\displaystyle
\frac{1}{\sqrt{2J\!+\!1}} \; \left\langle L,S_e,I,F,J \| S_e \| L,S_e,I,F,J \right\rangle \; \left\langle J 1 M_J 0 | J M_J \right\rangle \nonumber \\[3mm]
&=&\displaystyle
\frac{M_J} {\sqrt{J(J\!+\!1)(2J\!+\!1)}} \; \left\langle L,S_e,I,F,J \| S_e \| L,S_e,I,F,J
\right\rangle.
\end{array}
\end{equation}
The contribution to the $g$-factor of the hyperfine level under study is then:
\begin{equation}
g_1 (L,F,J) = g_e \; \frac{ \left\langle L,S_e,I,F,J \| S_e \| L,S_e,I,F,J \right\rangle} {\sqrt{J(J+1)(2J+1)}} \label{maingfactor}
\end{equation}
We now consider separately the cases of even and odd $L$.

\subsection{Even values of $L$}

In this case the intermediate angular momentum $F$ is irrelevant since $I=0$, and one directly has $\mathbf{J} = \mathbf{L} + \mathbf{S}_e$. The reduced matrix element appearing in~(\ref{maingfactor}) is then obtained as
\begin{equation}
\left\langle L S_e J \| S_e \| L S_e J \right\rangle = \left\langle S_e \| S_e \| S_e \right\rangle \; (-1)^{J+L+S_e+1} \; (2J+1) \; \left\{\begin{array}{ccc}
    S_e & 1 & S_e \\
     J  & L &  J
\end{array}\right\}
\end{equation}
where $\left\langle S_e \| S_e \| S_e \right\rangle =
\sqrt{S_e(S_e\!+\!1)(2S_e\!+\!1)}\> = \sqrt{{3}/{2}}\,$. One gets $g_1(L,J)$ for
the two possible values of $J$:
\begin{equation}
\begin{array}{rcl}
g_1(L,L\!+\!1/2) &=&\displaystyle ~\frac{g_e}{2L+1}, \\[3mm]
g_1(L,L\!-\!1/2) &=&\displaystyle - \frac{g_e}{2L+1}.
\end{array}
\end{equation}

\subsection{Odd values of $L$}

In this case the total spin of nuclei is one: $I=1$. The reduced matrix element appearing in~(\ref{maingfactor}) is obtained in two steps:
\begin{eqnarray}
\left\langle S_e,I,F \| S_e \| S_e,I,F \right\rangle &=& \left\langle S_e \| S_e \| S_e \right\rangle \; (-1)^{F+S_e+I+1} \; (2F+1) \;
\left\{\begin{array}{@{}ccc@{}}
    S_e & 1 & S_e \\
     F  & I &  F
\end{array}\right\}, \\
\left\langle L,S_e,I,F,J \| S_e \| L,S_e,I,F,J \right\rangle &=& \left\langle S_e,I,F \| S_e \| S_e,I,F \right\rangle \; (-1)^{J+L+F+1} \; (2J+1) \;
\left\{\begin{array}{@{}ccc@{}}
     F  & 1 &  F \\
     J  & L &  J
\end{array}\right\} \nonumber \\
= \sqrt{\frac{3}{2}} \; (-1)^{J+L+1/2} && (2F+1)\; (2J+1) \;
\left\{\begin{array}{@{}ccc@{}}
    S_e & 1 & S_e \\
     F  & I &  F
\end{array}\right\}
\left\{\begin{array}{@{}ccc@{}}
     F  & 1 &  F \\
     J  & L &  J
\end{array}\right\} \label{redmatelem}.
\end{eqnarray}
From this we get the factors $g_1 (L,F,J)$ for all hyperfine levels:
\begin{eqnarray}
\begin{array}{rcl}
g_1(L,1/2,L\!+\!1/2) &=&\displaystyle -\frac{g_e}{3}\>\frac{1}{2L+1} \\[3mm]
g_1(L,1/2,L\!-\!1/2) &=&\displaystyle ~\frac{g_e}{3}\>\frac{1}{2L+1} \\[3mm]
g_1(L,3/2,L\!+\!3/2) &=&\displaystyle ~g_e\>\frac{1}{2L+3} \\[3mm]
g_1(L,3/2,L\!+\!1/2) &=&\displaystyle ~\frac{g_e}{3} \> \frac{2L+9}{(2L\!+\!1)(2L\!+\!3)} \\[3mm]
g_1(L,3/2,L\!-\!1/2) &=&\displaystyle -\frac{g_e}{3} \> \frac{2L-7}{(2L\!-\!1)(2L\!+\!1)} \\[3mm]
g_1(L,3/2,L\!-\!3/2) &=&\displaystyle -g_e\>\frac{1}{2L-1}
\end{array}
\end{eqnarray}
\section{Contribution of the nuclear spin}
In the same way, we evaluate the contribution from the second term in equation~(\ref{hamiltonian}), which is nonzero only for odd values of $L$. Similarly to equation~(\ref{maingfactor}), the $g$-factor associated with this term is
\begin{equation}
g_2 (L,F,J) = -g_p \; \frac{m_e}{m_p} \; \frac{\left\langle L,S_e,I,F,J \| I \| L,S_e,I,F,J \right\rangle}{\sqrt{J(J+1)(2J+1)}}
\end{equation}
The reduced matrix element appearing in the above equation is obtained similarly to equation~(\ref{redmatelem}):
\begin{equation}
\left\langle L, S_e, I, F, J \| I \| L, S_e, I, F, J \right\rangle = \sqrt{6} \; (-1)^{J+L+1/2} \; (2F+1)(2J+1) \;
\left\{\begin{array}{@{}ccc@{}}
     I  & 1   &  I \\
     F  & S_e &  F
\end{array}\right\}
\left\{\begin{array}{@{}ccc@{}}
     F  & 1 &  F \\
     J  & L &  J
\end{array}\right\}.
\end{equation}
From this we deduce the factors $g_2(L,F,J)$ for all hyperfine levels:
\begin{equation}
\begin{array}{rcl}
g_2(L,1/2,L\!+\!1/2) &=&\displaystyle -g_p \; \frac{m_e}{m_p} \cdot \frac{4}{3}\> \frac{1}{2L+1}, \\[3mm]
g_2(L,1/2,L\!-\!1/2) &=&\displaystyle  ~g_p \; \frac{m_e}{m_p} \cdot \frac{4}{3}\> \frac{1}{2L+1}, \\[3mm]
g_2(L,3/2,L\!+\!3/2) &=&\displaystyle -g_p \; \frac{m_e}{m_p} \cdot \frac{2}{2L+3}, \\[3mm]
g_2(L,3/2,L\!+\!1/2) &=&\displaystyle -g_p \; \frac{m_e}{m_p} \cdot \frac{2}{3}\> \frac{2L+9}{(2L+1)(2L+3)}, \\[3mm]
g_2(L,3/2,L\!-\!1/2) &=&\displaystyle  ~g_p \; \frac{m_e}{m_p} \cdot \frac{2}{3}\> \frac{2L-7}{(2L-1)(2L+1)}, \\[3mm]
g_2(L,3/2,L\!-\!3/2) &=&\displaystyle  ~g_p \; \frac{m_e}{m_p} \cdot \frac{2}{2L-1}\>.
\end{array}
\end{equation}
\section{Contribution of the orbital momenta}
The contribution to the $g$-factor coming from the third and fourth terms in equation~(\ref{hamiltonian}) is
\begin{equation}
g_3 (v,L,F,J) = \frac{ \left\langle v,L,S_e,I,F,J \| L_e \| v,L,S_e,I,F,J \right\rangle} {\sqrt{J(J+1)(2J+1)}} - 2 \,\frac{m_e}{m_p} \> \frac{ \left\langle v,L,S_e,I,F,J \| L_1 \| v,L,S_e,I,F,J \right\rangle} {\sqrt{J(J+1)(2J+1)}}\>, \label{g3}
\end{equation}
where we have used that $\langle L_1 \rangle = \langle L_2 \rangle$ due to the symmetry of H$_2^+$ with respect to the exchange of nuclei. The reduced matrix elements appearing in the above expression are expressed as a function of reduced matrix elements involving only the orbital wave function:
\begin{equation}
\left\langle v, L, S_e, I, F, J \| L_i \| v, L, S_e, I, F, J \right\rangle = (-1)^{J+L+F+1} \; (2J+1) \; \left\{\begin{array}{ccc}
     L  & 1 &  L \\
     J  & F &  J
\end{array}\right\} \; \left\langle v, L \| L_i \| v, L \right\rangle.
\end{equation}
This expression is valid both for even and odd values of $L$ (in the first case $F\!=\!S_e\!=\!1/2$). We finally obtain the factors $g_3(v,L,F,J)$ for all hyperfine levels, expressed as a function of the orbital reduced matrix elements. For $F\!=\!1/2$ levels (both with even and odd $L$) we have:
\begin{equation}
\begin{array}{rcl}
g_3 (v,L,1/2,L\!+\!1/2) &=&\displaystyle \frac{2 \sqrt{L}}{\sqrt{L+1}\,(2L+1)} \; \langle\>\|L_{tot}\|\>\rangle \\[4mm]
g_3 (v,L,1/2,L\!-\!1/2) &=&\displaystyle \frac{2 \sqrt{L+1}}{\sqrt{L}\,(2L+1)} \; \langle\>\|L_{tot}\|\>\rangle
\end{array}
\end{equation}
and for $F=3/2$ levels (appearing only if $L$ is odd):
\begin{equation}
\begin{array}{rcl}
g_3 (v,L,3/2,L\!+\!3/2) &=&\displaystyle \frac{2 \sqrt{L}}{\sqrt{L+1}\, (2L+3)} \; \langle\>\|L_{tot}\|\>\rangle \\[3mm]
g_3 (v,L,3/2,L\!+\!1/2) &=&\displaystyle \frac{2 (2L^2 + 3L -3)}{\sqrt{L(L+1)}\,(2L+1)(2L+3)} \; \langle\>\|L_{tot}\|\>\rangle \\[3mm]
g_3 (v,L,3/2,L\!-\!1/2) &=&\displaystyle \frac{2 (2L^2 +  L -4)}{\sqrt{L(L+1)}\,(2L-1)(2L+1)} \; \langle\>\|L_{tot}\|\>\rangle \\[3mm]
g_3 (v,L,3/2,L\!-\!3/2) &=&\displaystyle \frac{2 \sqrt{L+1}}{\sqrt{L}\,(2L-1)} \; \langle\>\|L_{tot}\|\>\rangle
\end{array}
\end{equation}
where
\begin{equation}
\langle\>\|L_{tot}\|\>\rangle = \frac{\left\langle v, L \| L_e \| v, L \right\rangle} {\sqrt{2L+1}} - 2 \,\frac{m_e}{m_p} \> \frac{\left\langle v, L \| L_1 \| v, L \right\rangle} {\sqrt{2L+1}}. \label{ltot}
\end{equation}
The orbital matrix elements $\left\langle v, L \| L_e \| v, L \right\rangle$ and $\left\langle v, L \| L_1 \| v, L \right\rangle$ have been calculated using the variational approach presented in~\cite{paper1}, for $0 \leq v,L \leq 4$. We used basis lengths $N=$ 2000--3000, allowing a determination of the orbital momentum contribution $g_3$ with $10^{-4}$ relative accuracy. The results are shown in Table~\ref{orbmatelem}.
\begin{table}
\begin{tabular}{|@{\hspace{3mm}}c@{\hspace{3mm}}|@{\hspace{3mm}}c@{\hspace{3mm}}|@{\hspace{3mm}}c@{\hspace{3mm}}|@{\hspace{2mm}}c@{\hspace{2mm}}|@{\hspace{4mm}}c@{\hspace{4mm}}|@{\hspace{4mm}}c@{\hspace{3mm}}|}
\hline
\vrule width0pt height17pt depth10pt
$L$ & $v$ & $\displaystyle\frac{\langle v,L \| L_e \| v,L \rangle}{\sqrt{2L\!+\!1}}$   &  $\displaystyle\frac{\langle v,L \| L_1 \| v,L \rangle}{\sqrt{2L\!+\!1}}$  & $\langle\>\|L_{tot}\|\>\rangle$ & $g_{rot}$ \\
\hline
0 &   & 0 & 0 & 0 & \\
\hline
  & 0 & 0.615e$-$04 & 0.70708 & $-$0.7087e$-$03 & 0.9201 \\
  & 1 & 0.686e$-$04 & 0.70707 & $-$0.7015e$-$03 & 0.9108 \\
1 & 2 & 0.763e$-$04 & 0.70707 & $-$0.6938e$-$03 & 0.9008 \\
  & 3 & 0.847e$-$04 & 0.70706 & $-$0.6855e$-$03 & 0.8900 \\
  & 4 & 0.937e$-$04 & 0.70706 & $-$0.6764e$-$03 & 0.8782 \\
\hline
  & 0 & 1.069e$-$04 & 1.22469 & $-$1.2271e$-$03 & 0.9198 \\
  & 1 & 1.193e$-$04 & 1.22469 & $-$1.2146e$-$03 & 0.9105 \\
2 & 2 & 1.328e$-$04 & 1.22468 & $-$1.2012e$-$03 & 0.9004 \\
  & 3 & 1.473e$-$04 & 1.22467 & $-$1.1867e$-$03 & 0.8896 \\
  & 4 & 1.630e$-$04 & 1.22466 & $-$1.1710e$-$03 & 0.8778 \\
\hline
  & 0 & 1.521e$-$04 & 1.73197 & $-$1.7344e$-$03 & 0.9193 \\
  & 1 & 1.698e$-$04 & 1.73197 & $-$1.7167e$-$03 & 0.9100 \\
3 & 2 & 1.889e$-$04 & 1.73196 & $-$1.6977e$-$03 & 0.8998 \\
  & 3 & 2.095e$-$04 & 1.73195 & $-$1.6776e$-$03 & 0.8889 \\
  & 4 & 2.318e$-$04 & 1.73193 & $-$1.6547e$-$03 & 0.8771 \\
\hline
  & 0 & 1.980e$-$04 & 2.23597 & $-$2.2375e$-$03 & 0.9187 \\
  & 1 & 2.209e$-$04 & 2.23596 & $-$2.2146e$-$03 & 0.9092 \\
4 & 2 & 2.457e$-$04 & 2.23595 & $-$2.1898e$-$03 & 0.8991 \\
  & 3 & 2.725e$-$04 & 2.23593 & $-$2.1629e$-$03 & 0.8881 \\
  & 4 & 3.015e$-$04 & 2.23592 & $-$2.1339e$-$03 & 0.8761 \\
\hline
\end{tabular}
\caption{Reduced matrix elements of $\mathbf{L}_e$ and $\mathbf{L}_1$ (divided by $\sqrt{2L\!+\!1}$) for all ro-vibrational levels $(v,L)$ of H$_2^+$ with $0 \leq v,L \leq 4$. The deduced value of $\langle\>\|L_{tot}\|\>\rangle$ (defined in equation~(\ref{ltot})) is indicated in the third column. The last column contains the values of the rotational $g$-factors deduced from equation~(\ref{grot}). \label{orbmatelem}}
\end{table}

\section{Rotational $g$-factors in the strong-field regime}

These results allow a precise calculation of the rotational $g$-factor, which can be measured independently in a strong magnetic field. In this regime, the different angular momentum and spin vectors are decoupled, and the Zeeman Hamiltonian is usually written in the form
\begin{equation}
H_Z = g_e \mu_B \; \mathbf{S}_e \cdot \mathbf{B} - g_p \mu_p \; \mathbf{I} \cdot \mathbf{B}- g_{rot} \mu_p \; \mathbf{L} \cdot \mathbf{B}
\end{equation}
Comparing this expression with equation~(\ref{hamiltonian}), one easily obtains the expression of the rotational $g$-factor for a ro-vibrational level $(v,L)$:
\begin{equation}
g_{rot} (v,L) = \frac{1}{\sqrt{L(L+1)(2L+1)}} \; \left( 2 \left\langle v,L \| L_{1} \| v,L \right\rangle - \frac{m_p}{m_e} \; \left\langle v,L \| L_{e} \| v,L \right\rangle \right) \label{grot} = - \frac{m_p}{m_e} \; \frac{\langle\>\|L_{tot}\|\>\rangle}{\sqrt{L(L+1)}}
\end{equation}
The values of rotational $g$-factors are given in the last column of Table~\ref{orbmatelem}. They are improved with respect to the previous calculation performed within the adiabatic approximation by Rebane and Zotev \cite{rebane1997}. Note that the slight dependence of $g_{rot}$ on $L$ is neglected in the approach followed in that paper.

Loch et al.~\cite{loch1988} measured the rotational $g$-factor of H$_2^+$, averaged over the vibrational levels $v\!=\!4\!-\!6$ and the rotational levels $L\!=\!1\!-\!3$, to $g_{rot} = 0.920(40)$. Extending our calculations to the $v\!=\!5\!-\!6$ levels, and assuming the vibrational state populations reported in~\cite{loch1988}, we find $g_{rot} = 0.8688$, in disagreement with the experimental value by 1.28 $\sigma$.

\section{Effect of the state mixing} \label{section-mixing}

We now have to take into account the fact that for odd $L$, some of the hyperfine eigenstates are not pure states with a well-defined value of $F$, but linear combinations of $F=1/2$ and $F=3/2$ states (see equation~(\ref{eq-coefCi1})). The various contributions to the $g$-factor (denoted $\tilde{g}_J$ from now on) are changed in the following way:
\begin{equation}
\tilde{g}_J(v,L,\tilde{F},J) = \tilde{g}_1(v,L,\tilde{F},J) + \tilde{g}_2(v,L,\tilde{F},J) + \tilde{g}_3(v,L,\tilde{F},J),
\end{equation}
with
\begin{equation}
\begin{array}{rcl}
\tilde{g}_1(v,L,\tilde{F},L\!\pm\!1/2) &=& (C_1^{\pm})^2\, g_1(L,1/2,L\!\pm\!1/2) + (C_3^{\pm})^2\, g_1(L,3/2,L\!\pm\!1/2) \label{mixing1}\\[2mm]
&&\displaystyle +\>2\, C_1^{\pm}\, C_3^{\pm}\> g_e\> \frac{\left\langle L,S_e,I,1/2,L\!\pm\!1/2 \| S_e \| L,S_e,I,3/2,L\!\pm\!1/2 \right\rangle}{\sqrt{J(J+1)(2J+1)}},\\[3.5mm]
\tilde{g}_2(v,L,\tilde{F},L\!\pm\!1/2) &=& (C_1^{\pm})^2\, g_2(L,1/2,L\!\pm\!1/2) + (C_3^{\pm})^2\, g_2(L,3/2,L\!\pm\!1/2) \\[2mm]
&&\displaystyle -\>2\, C_1^{\pm}\, C_3^{\pm}\, g_p\> \frac{m_e}{m_p}\> \frac{\left\langle L,S_e,I,1/2,L\!\pm\!1/2 \| I \| L,S_e,I,3/2,L\!\pm\!1/2 \right\rangle}{\sqrt{J(J+1)(2J+1)}}, \\[3.5mm]
\tilde{g}_3(v,L,\tilde{F},L \!\pm\! 1/2) &=& (C_1^{\pm})^2\, g_3(v,L,1/2,L \!\pm\! 1/2) + (C_3^{\pm})^2 g_3(v,L,3/2,L \!\pm\! 1/2). \label{mixing3}
\end{array}
\end{equation}
The contribution $g_3$ does not contain any crossed terms, because the corresponding Hamiltonian acts only on orbital variables, and cannot couple $F=1/2$ states to $F=3/2$ states. The mixing coefficients $C_1^{\pm}$, $C_3^{\pm}$ are given in \cite{suppl}. It only remains to calculate the crossed reduced matrix elements appearing in the above expression. We have
\begin{equation}
\begin{array}{rcl}
\left\langle L,S_e,I,F,J \| S_e \| L,S_e,I,F',J \right\rangle &=& \alpha \; (-1)^{J+L+2F'+3/2}
\left\{\begin{array}{@{}ccc@{}}
   S_e  & 1 &  S_e \\
     F' & I &  F
\end{array}\right\} \left\langle S_e \| S_e \| S_e \right\rangle \\[3.5mm]
\left\langle L,S_e,I,F,J \|\, I\, \| L,S_e,I,F',J \right\rangle &=& \alpha \; (-1)^{J+L+3/2}
\left\{\begin{array}{@{}ccc@{}}
     I  & 1 &  I \\
     F' & S_e &  F
\end{array}\right\} \left\langle I \| \,I\, \| I \right\rangle,
\end{array}
\end{equation}
with
\begin{equation}
\alpha = \sqrt{\strut (2F+1)(2F'+1)}\>  (2J+1)
\left\{\begin{array}{@{}ccc@{}}
     F  & 1 &  F' \\
     J  & L &  J
\end{array}\right\}.
\end{equation}
From these expressions one may finally obtain
\begin{equation}
\begin{array}{rcl}
\left\langle L,1/2,1,1/2,L\!\pm\!1/2 \| S_e \| L,1/2,1,3/2,L\!\pm\!1/2 \right\rangle &=&\displaystyle \frac{\sqrt{8}}{3} \; \frac{\sqrt{L(L+1)}} {\sqrt{2L+1}} \\
\left\langle L,1/2,1,1/2,L\!\pm\!1/2 \|\, I\, \| L,1/2,1,3/2,L\!\pm\!1/2 \right\rangle &=&\displaystyle - \frac{\sqrt{8}}{3} \; \frac{\sqrt{L(L+1)}} {\sqrt{2L+1}}.
\end{array}
\end{equation}

\section{Final results and discussion}

\begin{table}
\begin{tabular}{|@{\hspace{3mm}}c@{\hspace{3mm}}|@{\hspace{3mm}}c@{\hspace{3mm}}|@{\hspace{3mm}}c@{\hspace{3mm}}|@{\hspace{3mm}}c@{\hspace{3mm}}|@{\hspace{3mm}}c@{\hspace{3mm}}|@{\hspace{3mm}}r@{\hspace{3mm}}|}
\hline
$L$ & $v$ & $J$ & $g_1(L,J)/g_e$  & $g_3(v,L,J)$  & $g_J(v,L,J)~$ \\
\hline
0 &   & 1/2 & $+$1   &     0       &  2.0023193  \\
\hline
  & 0 & 3/2 & $-$1/5 & $-$6.011e$-$04  & $-$0.4010650  \\
  &   & 5/2 & $+$1/5 & $-$4.008e$-$04  &    0.4000631  \\
  & 1 & 3/2 & $-$1/5 & $-$5.950e$-$04  & $-$0.4010589  \\
  &   & 5/2 & $+$1/5 & $-$3.967e$-$04  &    0.4000672  \\
2 & 2 & 3/2 & $-$1/5 & $-$5.885e$-$04  & $-$0.4010523  \\
  &   & 5/2 & $+$1/5 & $-$3.923e$-$04  &    0.4000716  \\
  & 3 & 3/2 & $-$1/5 & $-$5.814e$-$04  & $-$0.4010452  \\
  &   & 5/2 & $+$1/5 & $-$3.876e$-$04  &    0.4000763  \\
  & 4 & 3/2 & $-$1/5 & $-$5.737e$-$04  & $-$0.4010375  \\
  &   & 5/2 & $+$1/5 & $-$3.824e$-$04  &    0.4000814  \\
\hline
  & 0 & 7/2 & $-$1/9 & $-$5.559e$-$04  & $-$0.2230358  \\
  &   & 9/2 & $+$1/9 & $-$4.447e$-$04  &    0.2220352  \\
  & 1 & 7/2 & $-$1/9 & $-$5.502e$-$04  & $-$0.2230301  \\
  &   & 9/2 & $+$1/9 & $-$4.402e$-$04  &    0.2220398  \\
4 & 2 & 7/2 & $-$1/9 & $-$5.441e$-$04  & $-$0.2230240  \\
  &   & 9/2 & $+$1/9 & $-$4.352e$-$04  &    0.2220447  \\
  & 3 & 7/2 & $-$1/9 & $-$5.374e$-$04  & $-$0.2230173  \\
  &   & 9/2 & $+$1/9 & $-$4.299e$-$04  &    0.2220500  \\
  & 4 & 7/2 & $-$1/9 & $-$5.302e$-$04  & $-$0.2230101  \\
  &   & 9/2 & $+$1/9 & $-$4.241e$-$04  &    0.2220558  \\
\hline
\end{tabular}
\caption{$g$-factors of all hyperfine states for the ro-vibrational levels $(v,L)$ with $L=0,2,4$ and $0 \leq v \leq 4$. The fourth column is the contribution from the electron spin magnetic moment, the fifth one is the contribution from orbital momenta, and the last one is the total value of $g_J(v,L,J)$. All digits are converged. The relative theoretical accuracy on all contributions (and on $g_J$) is $\mathcal{O}(\alpha^2)$. \label{final-evenL}}
\end{table}

\begin{table}
\tiny
\begin{tabular}{|@{\hspace{2mm}}c@{\hspace{2mm}}|@{\hspace{2mm}}c@{\hspace{2mm}}|@{\hspace{2mm}}c@{\hspace{2mm}}|@{\hspace{2mm}}c@{\hspace{2mm}}|@{\hspace{2mm}}c@{\hspace{2mm}}|@{\hspace{2mm}}c@{\hspace{2mm}}|@{\hspace{2mm}}c@{\hspace{2mm}}|@{\hspace{2mm}}c@{\hspace{2mm}}|@{\hspace{2mm}}c@{\hspace{2mm}}|@{\hspace{2mm}}c@{\hspace{2mm}}|@{\hspace{2mm}}c@{\hspace{2mm}}|}
\hline
\vrule width0pt height15pt depth10pt
$v$ & $\tilde{F}$ & $J$ & $g_1/g_e$  & $\displaystyle\frac{m_p}{m_e}\,\frac{g_2}{g_p}$ & $g_3 $  & $g_J(v,L,\tilde{F},J)$ & $\tilde{g}_1/g_e$  & $\displaystyle\frac{m_p}{m_e}\,\frac{\tilde{g}_2}{g_p}$ & $\tilde{g}_3 $  & $\tilde{g}_J(v,L,\tilde{F},J)$ \\
\hline
  & 1/2 & 1/2 & $+$1/9   & $+$4/9   & $-$6.682e$-$04 & $+$0.2231638 & $+$0.0629313 & $+$0.3932396 & $-$6.667e$-$04 & $+$0.1265381 \\
  &     & 3/2 & $-$1/9   & $-$4/9   & $-$3.341e$-$04 & $-$0.2241660 & $-$0.1172298 & $-$0.4506607 & $-$3.340e$-$04 & $-$0.2364365 \\
0 & 3/2 & 1/2 & $+$5/9   & $-$10/9  & $+$3.341e$-$04 & $+$1.1093536 & $+$0.6037354 & $-$1.0599063 & $+$3.326e$-$04 & $+$1.2059793 \\
  &     & 3/2 & $+$11/45 & $-$22/45 & $-$1.336e$-$04 & $+$0.4878350 & $+$0.2505632 & $-$0.4826727 & $-$1.337e$-$04 & $+$0.5001054 \\
  &     & 5/2 & $+$1/5   & $-$2/5   & $-$2.005e$-$04 & $+$0.3990466 &              &              &                & $+$0.3990466 \\
\hline
  & 1/2 & 1/2 & $+$1/9   & $+$4/9   & $-$6.614e$-$04 & $+$0.2231705 & $+$0.0648181 & $+$0.3953621 & $-$6.600e$-$04 & $+$0.1303292 \\
  &     & 3/2 & $-$1/9   & $-$4/9   & $-$3.307e$-$04 & $-$0.2241627 & $-$0.1170219 & $-$0.4504461 & $-$3.307e$-$04 & $-$0.2360162 \\
1 & 3/2 & 1/2 & $+$5/9   & $-$10/9  & $+$3.307e$-$04 & $+$1.1093503 & $+$0.6018486 & $-$1.0620288 & $+$3.293e$-$04 & $+$1.2021916 \\
  &     & 3/2 & $+$11/45 & $-$22/45 & $-$1.323e$-$04 & $+$0.4878363 & $+$0.2503552 & $-$0.4828872 & $-$1.323e$-$04 & $+$0.4996898 \\
  &     & 5/2 & $+$1/5   & $-$2/5   & $-$1.984e$-$04 & $+$0.3990486 &              &              &                & $+$0.3990486 \\
\hline
  & 1/2 & 1/2 & $+$1/9   & $+$4/9   & $-$6.541e$-$04 & $+$0.2231778 & $+$0.0667416 & $+$0.3975158 & $-$6.529e$-$04 & $+$0.1341944 \\
  &     & 3/2 & $-$1/9   & $-$4/9   & $-$3.271e$-$04 & $-$0.2241590 & $-$0.1168072 & $-$0.4502248 & $-$3.270e$-$04 & $-$0.2355819 \\
2 & 3/2 & 1/2 & $+$5/9   & $-$10/9  & $+$3.271e$-$04 & $+$1.1093466 & $+$0.5999250 & $-$1.0641825 & $+$3.258e$-$04 & $+$1.1983300 \\
  &     & 3/2 & $+$11/45 & $-$22/45 & $-$1.308e$-$04 & $+$0.4878378 & $+$0.2501405 & $-$0.4831085 & $-$1.309e$-$04 & $+$0.4992607 \\
  &     & 5/2 & $+$1/5   & $-$2/5   & $-$1.962e$-$04 & $+$0.3990508 &              &              &                & $+$0.3990508 \\
\hline
  & 1/2 & 1/2 & $+$1/9   & $+$4/9   & $-$6.463e$-$04 & $+$0.2231857 & $+$0.0687069 & $+$0.3997057 & $-$6.451e$-$04 & $+$0.1381439 \\
  &     & 3/2 & $-$1/9   & $-$4/9   & $-$3.231e$-$04 & $-$0.2241551 & $-$0.1165853 & $-$0.4499964 & $-$3.231e$-$04 & $-$0.2351330 \\
3 & 3/2 & 1/2 & $+$5/9   & $-$10/9  & $+$3.231e$-$04 & $+$1.1093427 & $+$0.5979598 & $-$1.0663724 & $+$3.220e$-$04 & $+$1.1943845 \\
  &     & 3/2 & $+$11/45 & $-$22/45 & $-$1.293e$-$04 & $+$0.4878393 & $+$0.2499186 & $-$0.4833369 & $-$1.293e$-$04 & $+$0.4988172 \\
  &     & 5/2 & $+$1/5   & $-$2/5   & $-$1.939e$-$04 & $+$0.3990532 &              &              &                & $+$0.3990532 \\
\hline
  & 1/2 & 1/2 & $+$1/9   & $+$4/9   & $-$6.377e$-$04 & $+$0.2231942 & $+$0.0707213 & $+$0.4019394 & $-$6.367e$-$04 & $+$0.1421927 \\
  &     & 3/2 & $-$1/9   & $-$4/9   & $-$3.189e$-$04 & $-$0.2241508 & $-$0.1163546 & $-$0.4497592 & $-$3.188e$-$04 & $-$0.2346661 \\
4 & 3/2 & 1/2 & $+$5/9   & $-$10/9  & $+$3.189e$-$04 & $+$1.1093384 & $+$0.5959453 & $-$1.0686061 & $+$3.179e$-$04 & $+$1.1903399 \\
  &     & 3/2 & $+$11/45 & $-$22/45 & $-$1.275e$-$04 & $+$0.4878410 & $+$0.2496879 & $-$0.4835741 & $-$1.276e$-$04 & $+$0.4983563 \\
  &     & 5/2 & $+$1/5   & $-$2/5   & $-$1.913e$-$04 & $+$0.3990557 &              &              &                & $+$0.3990557 \\
\hline
\end{tabular}
\caption{$g$-factors of all hyperfine states for the ro-vibrational levels $(v,L)$ with $L=1$ and $0 \leq v \leq 4$. The fourth column is the contribution from the electron spin magnetic moment, the fifth one is the contribution from the nuclear spin magnetic moment, and the sixth one is the contribution from orbital momenta. All three terms are calculated without taking state mixing into account. The seventh column is the sum of these contributions. Columns 8 to 10 are the corrected values of the three contributions obtained by taking state mixing into account. The last column is the final value of the $g$-factor. All digits are converged. The relative theoretical accuracy for all contributions is $\mathcal{O}(\alpha^2)$. \label{final-L1}}
\end{table}

\begin{table}
\tiny
\begin{tabular}{|@{\hspace{2mm}}c@{\hspace{2mm}}|@{\hspace{2mm}}c@{\hspace{2mm}}|@{\hspace{2mm}}c@{\hspace{2mm}}|@{\hspace{2mm}}c@{\hspace{2mm}}|@{\hspace{2mm}}c@{\hspace{2mm}}|@{\hspace{2mm}}c@{\hspace{2mm}}|@{\hspace{2mm}}c@{\hspace{2mm}}|@{\hspace{2mm}}c@{\hspace{2mm}}|@{\hspace{2mm}}c@{\hspace{2mm}}|@{\hspace{2mm}}c@{\hspace{2mm}}|@{\hspace{2mm}}c@{\hspace{2mm}}|}
\hline
\vrule width0pt height15pt depth10pt
$v$ & $\tilde{F}$ & $J$ & $g_1/g_e$  & $\displaystyle\frac{m_p}{m_e}\>\frac{g_2}{g_p}$ & $g_3 $  & $g_J(v,L,\tilde{F},J)$ & $\tilde{g}_1/g_e$  & $\displaystyle\frac{m_p}{m_e}\>\frac{\tilde{g}_2}{g_p}$ & $\tilde{g}_3 $  & $\tilde{g}_J(v,L,\tilde{F},J)$ \\
\hline
  & 1/2 & 5/2 & $+$1/21  & $+$4/21  & $-$5.722e$-$04 & $+$0.0953558 & $+$0.0264523 & $+$0.1686545 & $-$5.719e$-$04 & $+$0.0529071 \\
  &     & 7/2 & $-$1/21  & $-$4/21  & $-$4.292e$-$04 & $-$0.0963571 & $-$0.0566485 & $-$0.1996746 & $-$4.291e$-$04 & $-$0.1144649 \\
0 & 3/2 & 3/2 & $-$1/5   & $+$2/5   & $-$8.011e$-$04 & $-$0.4000481 &              &              &                & $-$0.4000481 \\
  &     & 5/2 & $+$1/105 & $-$2/105 & $-$4.864e$-$04 & $+$0.0185254 & $+$0.0306905 & $+$0.0027741 & $-$4.867e$-$04 & $+$0.0609740 \\
  &     & 7/2 & $+$5/63  & $-$10/63 & $-$3.815e$-$04 & $+$0.1580499 & $+$0.0883945 & $-$0.1495318 & $-$3.816e$-$04 & $+$0.1761576 \\
  &     & 9/2 & $+$1/9   & $-$2/9   & $-$3.338e$-$04 & $+$0.2214701 &              &              &                & $+$0.2214701 \\
\hline
  & 1/2 & 5/2 & $+$1/21  & $+$4/21  & $-$5.664e$-$04 & $+$0.0953616 & $+$0.0272667 & $+$0.1695181 & $-$5.661e$-$04 & $+$0.0545462 \\
  &     & 7/2 & $-$1/21  & $-$4/21  & $-$4.248e$-$04 & $-$0.0963528 & $-$0.0563438 & $-$0.1993583 & $-$4.247e$-$04 & $-$0.1138494 \\
1 & 3/2 & 3/2 & $-$1/5   & $+$2/5   & $-$7.929e$-$04 & $-$0.4000400 &              &              &                & $-$0.4000403 \\
  &     & 5/2 & $+$1/105 & $-$2/105 & $-$4.814e$-$04 & $+$0.0185303 & $+$0.0298762 & $+$0.0019104 & $-$4.817e$-$04 & $+$0.0593457 \\
  &     & 7/2 & $+$5/63  & $-$10/63 & $-$3.776e$-$04 & $+$0.1580538 & $+$0.0880898 & $-$0.1498480 & $-$3.777e$-$04 & $+$0.1755504 \\
  &     & 9/2 & $+$1/9   & $-$2/9   & $-$3.304e$-$04 & $+$0.2214735 &              &              &                & $+$0.2214735 \\
\hline
  & 1/2 & 5/2 & $+$1/21  & $+$4/21  & $-$5.601e$-$04 & $+$0.0953679 & $+$0.0280999 & $+$0.1703997 & $-$5.598e$-$04 & $+$0.0562234 \\
  &     & 7/2 & $-$1/21  & $-$4/21  & $-$4.201e$-$04 & $-$0.0963480 & $-$0.0560290 & $-$0.1990321 & $-$4.200e$-$04 & $-$0.1132134 \\
2 & 3/2 & 3/2 & $-$1/5   & $+$2/5   & $-$7.841e$-$04 & $-$0.4000311 &              &              &                & $-$0.4000313 \\
  &     & 5/2 & $+$1/105 & $-$2/105 & $-$4.761e$-$04 & $+$0.0185357 & $+$0.0290430 & $+$0.0010288 & $-$4.763e$-$04 & $+$0.0576802 \\
  &     & 7/2 & $+$5/63  & $-$10/63 & $-$3.734e$-$04 & $+$0.1580580 & $+$0.0877750 & $-$0.1501742 & $-$3.735e$-$04 & $+$0.1749233 \\
  &     & 9/2 & $+$1/9   & $-$2/9   & $-$3.267e$-$04 & $+$0.2214772 &              &              &                & $+$0.2214772 \\
\hline
  & 1/2 & 5/2 & $+$1/21  & $+$4/21  & $-$5.533e$-$04 & $+$0.0953747 & $+$0.0289535 & $+$0.1713009 & $-$5.530e$-$04 & $+$0.0579423 \\
  &     & 7/2 & $-$1/21  & $-$4/21  & $-$4.150e$-$04 & $-$0.0963429 & $-$0.0557027 & $-$0.1986944 & $-$4.149e$-$04 & $-$0.1125539 \\
3 & 3/2 & 3/2 & $-$1/5   & $+$2/5   & $-$7.746e$-$04 & $-$0.4000216 &              &              &                & $-$0.4000217 \\
  &     & 5/2 & $+$1/105 & $-$2/105 & $-$4.703e$-$04 & $+$0.0185415 & $+$0.0281893 & $+$0.0001276 & $-$4.705e$-$04 & $+$0.0559739 \\
  &     & 7/2 & $+$5/63  & $-$10/63 & $-$3.689e$-$04 & $+$0.1580625 & $+$0.0874487 & $-$0.1505119 & $-$3.689e$-$04 & $+$0.1742735 \\
  &     & 9/2 & $+$1/9   & $-$2/9   & $-$3.227e$-$04 & $+$0.2214812 &              &              &                & $+$0.2214812 \\
\hline
  & 1/2 & 5/2 & $+$1/21  & $+$4/21  & $-$5.459e$-$04 & $+$0.0953821 & $+$0.0298308 & $+$0.1722248 & $-$5.457e$-$04 & $+$0.0597089 \\
  &     & 7/2 & $-$1/21  & $-$4/21  & $-$4.094e$-$04 & $-$0.0963374 & $-$0.0553637 & $-$0.1983442 & $-$4.094e$-$04 & $-$0.1118686 \\
4 & 3/2 & 3/2 & $-$1/5   & $+$2/5   & $-$7.643e$-$04 & $-$0.4000113 &              &              &                & $-$0.4000115 \\
  &     & 5/2 & $+$1/105 & $-$2/105 & $-$4.640e$-$04 & $+$0.0185477 & $+$0.0273121 & $-$0.0007963 & $-$4.642e$-$04 & $+$0.0542209 \\
  &     & 7/2 & $+$5/63  & $-$10/63 & $-$3.639e$-$04 & $+$0.1580674 & $+$0.0871097 & $-$0.1508622 & $-$3.640e$-$04 & $+$0.1735986 \\
  &     & 9/2 & $+$1/9   & $-$2/9   & $-$3.184e$-$04 & $+$0.2214855 &              &              &                & $+$0.2214855 \\
\hline
\end{tabular}
\caption{Same as Table~\ref{final-L1}, for the rotational level $L=3$. \label{final-L3}}
\end{table}

Our results are summarized in Tables~\ref{final-evenL}, \ref{final-L1} and \ref{final-L3}, where all contributions are summed up in order to obtain the $g$-factors for all hyperfine levels. The accuracy of this calculation can be limited by several factors.

The first one is the variational calculation of the orbital momentum reduced matrix elements appearing in expression~(\ref{g3}) of $g_3$, the relative accuracy of which is $10^{-4}$. This results in a numerical uncertainty of less than $10^{-7}$ on the final $g$-factor values.

The second limitation comes from the coefficients of the hyperfine Hamiltonian, which are known with a relative accuracy $\mathcal{O}(\alpha^2) \!\sim\! 5\!\times\!10^{-5}$. The mixing coefficients defined in~(\ref{eq-coefCi1}) are affected by this uncertainty, resulting in an uncertainty comprised between $5\!\times\! 10^{-7}$ and $5\!\times\!10^{-6}$ for the final values of $\tilde{g}(L,\tilde{F},J)$, depending on the magnitude of the corrections due to state mixing. This uncertainty does not affect the hyperfine states which are pure states (i.e. all states of even $L$, and states of odd $L$ with $F = 3/2$ and $J = L\!\pm\!3/2$).

Finally, and most importantly, relativistic and radiative corrections, as considered by Hegstrom in the strong-field regime in \cite{hegstrom1979}, are not included. This limits the relative accuracy to $\mathcal{O}(\alpha^2) \!\sim\! 5\!\times\!10^{-5}$.

To sum it up, the accuracy of our results is $\mathcal{O}(\alpha^2) \!\sim\! 5\!\times\!10^{-5}$. Inclusion of the leading-order QED corrections would improve it to $\mathcal{O}(\alpha^3) \!\sim\! 3\!\times\!10^{-7}$. Such accuracy would appear to be hard to meet in rf experiments with weak magnetic fields. $g$-factor measurements on H$_2^+$ with an accuracy in the ppm range were achieved in strong magnetic fields, using spin-dependent charge-exchange techniques~\cite{loch1988}.

We have nevertheless given the $g$-factors values with 7 digits. Although all digits are not significant, this is helpful for understanding the order of magnitude of various effects, such as the variation as a function of $v$ and $L$, or the importance of corrections due to state mixing. For example, it can be seen that state mixing acts on the value of the orbital contribution $g_3$ at the level of $10^{-6}$ at most, so this effect may be neglected at the present level of accuracy.

To our knowledge, the magnetic moments of the H$_2^+$ hyperfine states in a weak magnetic field have been investigated only by Richardson, Jefferts and Dehmelt in 1968~\cite{richardson1968}. They give a few ratios of $g$-factors between different hyperfine states, which we have reported in Table~\ref{gfactor-exp} together with the result of our calculation. Note that the experimental values are averaged over the vibrational states $v > 4$. In our evaluation, we have taken the vibrational states $v =$ 5-8 into account, and assumed that their relative populations (determined by the creation process by electron impact ionization of H$_2$ at room temperature) are the same as measured in Ref.~\cite{weijun1993}. Good agreement is obtained in all cases.

\begin{table}[b]
\begin{tabular}{|@{\hspace{4mm}}c@{\hspace{4mm}}|@{\hspace{4mm}}c@{\hspace{4mm}}|@{\hspace{4mm}}c@{\hspace{4mm}}|}
\hline
\vrule width0pt height12pt depth4pt
Ratio & Calculated & Measured \cite{richardson1968}\\
\hline
\vrule width0pt height13pt depth4pt
$g_J(L\!=\!1,\tilde{F}\!=\!1/2,J\!=\!3/2)\,\big/\,g_J(L\!=\!1,F\!=\!3/2,J\!=\!5/2)$ & 0.5855 & 0.584(3) \\
\vrule width0pt height11pt depth4pt
$g_J(L\!=\!1,\tilde{F}\!=\!3/2,J\!=\!3/2)\,\big/\,g_J(L\!=\!1,F\!=\!3/2,J\!=\!5/2)$ & 1.2463 & 1.241(6) \\
\vrule width0pt height11pt depth5pt
$g_J(L\!=\!1,\tilde{F}\!=\!1/2,J\!=\!3/2)\,\big/\,g_J(L\!=\!3,F\!=\!3/2,J\!=\!9/2)$ & 1.0549 & 1.051(5) \\
\hline
\end{tabular}
\caption{$g$-factor ratios of some hyperfine states of H$_2^+$. \label{gfactor-exp}}
\end{table}

We now use these results to evaluate the Zeeman shift and splitting of the two-photon transitions $(v\!=\!0,L) \rightarrow (v'\!=\!1,L)$ envisaged for high-precision spectroscopy of the H$_2^+$ ion. For illustration, we choose a magnetic field of the order of the earth field, $B = 5\!\times\! 10^{-5}$~T. As explained in the introduction, the accuracy of our calculation is more than sufficient for this purpose. The leading relativistic corrections which we have neglected correspond to a shift of order $\alpha^2 \mu_B B \sim$ 35 Hz, well below the present goal accuracy of spectroscopy experiments. The frequency shift $\Delta \nu$ of a two-photon transition $(v,L,\tilde{F},J) \rightarrow (v',L,\tilde{F}',J')$ in the magnetic field $B$ is
\begin{equation}
2 h \Delta \nu  = \left[ M'_J \tilde{g}_J (v',L,\tilde{F}',J') - M_J \tilde{g}_J (v,L,\tilde{F},J) \right] \mu_B B \\
\end{equation}
If circular polarization is used, the selection rule is $M'_J-M_J\!=\!2$. $M_J$ can take the possible values $-J,-J+1...J-2$ so that the shift of the line centre (corresponding to $M_J\!=\!-1$) is
\begin{equation}
2 h \Delta \nu = \left[ \tilde{g}_J (v,L,\tilde{F},J) + \tilde{g}_J (v',L,\tilde{F}',J') \right] \mu_B B \label{circular}
\end{equation}
In the case of linear polarization, the selection rule is $M'_J-M_J\!=\!0$, and $M_J$ can take the possible values $-J,-J+1...J$. There is no global shift, and the Zeeman splitting between extreme values of $M_J$ is
\begin{equation}
h \Delta \nu = \left[ \tilde{g}_J (v',L,\tilde{F}',J') - \tilde{g}_J (v,L,\tilde{F},J) \right] J \mu_B B \label{linear}
\end{equation}
We have shown in~\cite{paper1} that the most intense hyperfine components are those between pairs of homologous spin states, $(F,J) \rightarrow (F,J)$, and only these components are considered in the following. In this case, we benefit from an almost complete cancellation (to 1 percent or better) between the $g$-factors of the initial and final states, so that the Zeeman splitting is very small (compared to the global shift observed in circular polarization). Note that such cancellation will also take place for relativistic corrections, so that the theoretical uncertainty is also reduced. In estimating the uncertainties, we have assumed cancellation to 1 percent.
\begin{table}
\begin{tabular}{|@{\hspace{3mm}}c@{\hspace{3mm}}|@{\hspace{4mm}}c@{\hspace{4mm}}|@{\hspace{4mm}}c@{\hspace{4mm}}|@{\hspace{4mm}}c@{\hspace{4mm}}|@{\hspace{4mm}}c@{\hspace{4mm}}|}
\hline
\vrule width0pt height12pt depth4pt
$L$ & $F$ & $J$ & shift & splitting \\
    &     &     & ($\sigma_+$ polar.) & ($\pi$ polar.) \\
\hline
\vrule width0pt height11pt depth4pt
0  & 1/2 & 1/2 & * & 0 \\
\hline
\vrule width0pt height11pt depth0pt
   & 1/2 & 1/2 &      *         & 1327  \\
   &     & 3/2 & $-$165 314(9)  & 441   \\
1  & 3/2 & 1/2 &      *         & 1325  \\
   &     & 3/2 & $+$349 834(19) & 436   \\
\vrule width0pt height8pt depth4pt
   &     & 5/2 & $+$279 258(15) & 3.5(3) \\
\hline
\vrule width0pt height11pt depth0pt
2  & 1/2 & 3/2 & $-$280 668(15) & 6.4(2) \\
\vrule width0pt height8pt depth4pt
   & 1/2 & 5/2 & $+$279 971(15) & 7.2(3) \\
\hline
\vrule width0pt height11pt depth0pt
   & 1/2 & 5/2 & $+$ 37 599(2) & 2868  \\
   &     & 7/2 & $-$ 79 889(4) & 1508  \\
3  & 3/2 & 3/2 & $-$279 956(15)& 8.2(2)\\
   &     & 5/2 & $+$ 42 101(2) & 2849  \\
   &     & 7/2 & $+$123 065(7) & 1487  \\
\vrule width0pt height8pt depth4pt
   &     & 9/2 & $+$154 989(8) & 10.7(3)\\
\hline
\end{tabular}
\caption{Zeeman shift and splitting of two-photon transition lines $(v\!=\!0,L,F,J) \rightarrow (v'\!=\!1,L,F,J)$ in a magnetic fields of $5\!\times\!10^{-5}$ T, in Hz. The fourth column is the predicted shift of the line center in the case of $\sigma_+$ excitation polarizations (equation~(\ref{circular})). The stars indicate transitions which are forbidden in circular polarization. An estimate of the theoretical uncertainty (corresponding to a relative accuracy of $\mathcal{O}(\alpha^2)$) is given. The last column is the Zeeman splitting, i.e. the frequency difference between extreme values of $M_J$, evaluated in the linear polarization case (equation~(\ref{linear})). An estimate of the theoretical uncertainty (see text) is given when it is significant. \label{zeeman2photon}}
\end{table}

The cases $L=1,2,3$ are compared in Table~\ref{zeeman2photon} for the cases of circular and linear excitation polarizations. When circular polarization is used, the two-photon transition lines are typically shifted by a few hundreds of kHz. This does not represent in itself a limitation of experimental accuracy, since it is possible take the average of measurements in $\sigma_+$ and $\sigma_-$ polarizations. However, it also means that magnetic field fluctuations of the order of 10 mG result in a line broadening of order 1--10 kHz. If one wishes to improve the resolution beyond this limit, active control and stabilization of the magnetic field is required~\cite{ringot2001}. Together with larger transition probabilities as discussed in~\cite{paper1}, this brings a strong argument in favor of using linear polarizations. Unfortunately, optical isolation of the laser source from feedback by the enhancement cavity is at present only possible with a polariser followed by a quarter-wave plate, which imposes working with circular polarizations~\cite{paper1}. One solution is to add a transverse magnetic field, which must be sufficiently strong to separate the three components $\Delta M_J = 0, \pm 2$. It can be seen from Table~\ref{zeeman2photon} that a field in the $10^{-5}$--$10^{-4}$ T range (depending on the transition) is enough to obtain a separation of the order of 100 kHz, i.e. clearly resolved components assuming a linewidth of a few kHz~\cite{paper1}.

Finally, the Zeeman splitting is extremely small (a few Hz) when the states involved in the two-photon transition are pure states, while it is of a few kHz in other cases. This makes such transitions especially attractive from the metrological point of view.

\section{Conclusion}

We have obtained improved values of $g$-factors of the hyperfine states of the hydrogen molecular ion, which are in good agreement with experiment. The achieved accuracy is $\mathcal{O}(\alpha^2) \sim 5\!\times\!10^{-5}$. The accuracy on the rotational $g$-factors has also been improved by use of a variational method allowing to take the full three body dynamics into account. We have used these results to evaluate the Zeeman shift and splitting of several two-photon vibrational transition lines, and shown that transitions involving pure hyperfine states (i.e. all states of even $L$, and states of odd $L$ with $F =$ 3/2 and $J = L \pm$3/2) benefit from a very small Zeeman splitting.

This work was supported by l'Universit\'e D'Evry Val d'Essonne. V.I.K.\ acknowledges support of the Russian Foundation for Basic Research under Grant No.\ 08-02-00341. Laboratoire Kastler Brossel de l'Universit\'e Pierre et Marie Curie et de l'Ecole Normale Sup\'erieure is UMR 8552 du CNRS.

\end{document}